\documentclass[twocolumn,prl,aps,10pt,floatfix]{revtex4-1}
\usepackage{amssymb, amsmath, amsthm}	
\usepackage{graphicx}	
\usepackage{color}
\usepackage{units}

\usepackage{natbib}

\begin{document}

\title{Optical transmission through a dipolar layer}

\author{James Keaveney}
\author{Ulrich Krohn} 
\author{Julia Gontcharov} 
\author{Ifan G. Hughes}
\author{Charles S. Adams} 
\affiliation{Department of Physics, Rochester Building, Durham University, South Road, Durham DH1 3LE, United Kingdom}

\author{Armen Sargsyan}
\author{David Sarkisyan}
\affiliation{Institute for Physical Research, National Academy of Sciences - Ashtarak 2, 0203, Armenia}

\date{\today}

\maketitle

{\bf 
%
%
The interaction between light and matter is fundamental to developments in quantum optics and information. Over recent years enormous progress has been made in controlling the interface between light and single emitters including
ions \cite{Harlander2011},  
atoms \cite{Raimond2001,Alton2011}, 
molecules \cite{Celebrano2011},  
quantum dots  \cite{Shields2007},
and ensembles  \cite{Hammerer2010}. 
For many systems, inter-particle interactions are typically negligible. However, if the emitters are separated by less than the emission wavelength, $\lambda$, resonant dipole--dipole interactions modify the radiative decay rate \cite{DeVoe1996} and induce a splitting or shift of the resonance \cite{Hettich2002}. 
Here we map out the transition between individual dipoles and a strongly interacting ensemble by increasing the density of atoms confined in a layer with thickness much less than $\lambda$. We find two surprising results: 
whereas for a non-interacting ensemble the opacity increases linearly with atomic density, for an interacting ensemble the opacity saturates,
i.e., a thin dipolar layer never becomes opaque regardless of how many scatterers are added. 
Secondly, the relative phase of the dipoles produces an abrupt change in the optical transmission around the thickness} $\lambda/4$\textbf{.}

The requirement for strong dipole--dipole interactions --- that the atoms must separated by less than $\lambda$ --- places stringent demands on the spatial extent of the ensemble.  For atomic gases considerable progress has been made in the ability to confine atoms in thin layers \cite{Sarkisyan2001,Sarkisyan2004},
optical fibers \cite{Slepkov2010,Bajcsy2011} 
and waveguides \cite{Wu2010}. 
Alternatively, it has been recognised that the spatial constraints can be relaxed by exploiting the long wavelength dipoles associated with Rydberg transitions \cite{Lukin2001,Pritchard2011a} where strong dipole--dipole interactions between Rydberg atoms give rise to a large optical non--linearity \cite{Pritchard2010,Gorshkov2011}.
However, micron--scale spatial confinement is still required \cite{Kubler2010}. 
As in the two--level case, in Rydberg systems dephasing can also play a role in the emission process \cite{Honer2011,Bariani2011}.

\begin{figure*}[ht]
\includegraphics[width=0.84\textwidth,angle=0]{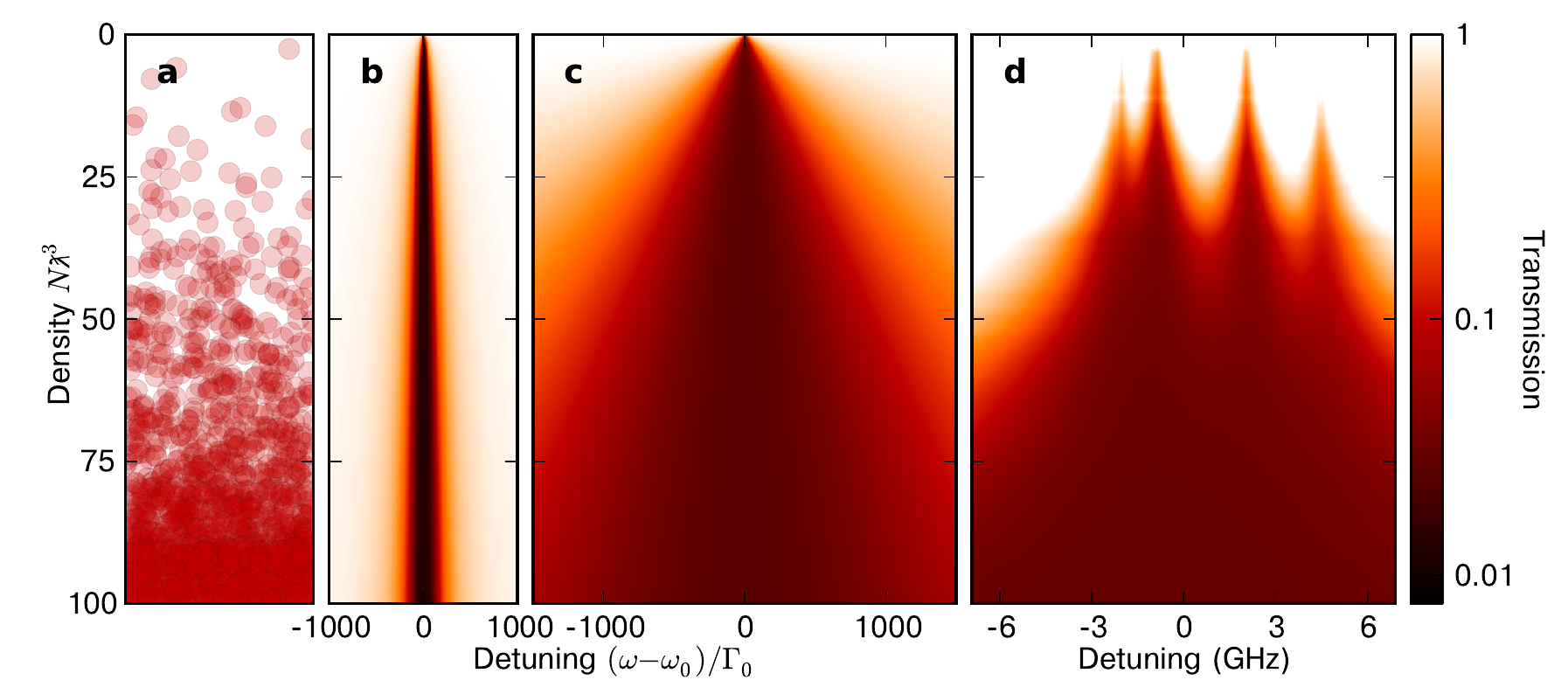}
\caption{Evolution of opacity with density.
Dipoles are represented in~(\textbf{a}) by semitransparent discs, whose radii are the dipole--dipole interaction range $\lambdabar$. At low density the discs do not overlap on average, and hence the interactions are negligible, whereas at high density the interactions dominate.
 (\textbf{b})~With no dipole--dipole interactions the lineshape is a Lorentzian with constant width and the sample quickly becomes optically thick (black). (\textbf{c})~When dipole--dipole interactions are considered, the line broadens linearly with density, eventually leading to a constant broadband value at high density. (\textbf{d})~Experimental data for cell thickness $\delta z=390$~nm showing transmission over the Rb D$_{2}$ line, where there are now four separate absorption lines owing to the hyperfine splitting of the ground states of $^{85}$Rb and $^{87}$Rb. Opacity is represented by the colour map, which goes from completely transparent (white) through to completely opaque (black).
}
 \label{fig:dd}
\end{figure*}

The underlying mechanism of light scattering is the interference between the incident field and the field produced by induced oscillatory dipoles. 
For independent dipoles, on resonance, the destructive interference in the forward direction is equivalent to removing a cross-sectional area of the incident beam equivalent to $\sigma_{0}=6\pi\lambdabar^2$ per atom, where $\lambdabar = \lambda/(2\pi)$ is the reduced wavelength of the dipolar radiation, corresponding to the effective length scale of the light-matter interaction. 
If the angular frequency $\omega$ of the incident field is detuned from the resonance frequency $\omega_{0}$ by $\Delta = \omega-\omega_{0}$, the optical cross section becomes
\begin{align}
\sigma = \frac{\sigma_{0}}{1+ 4(\Delta/\Gamma)^{2}}~,
\label{eq:OD}
\end{align}
where $\Gamma$ is the full width at half-maximum of the resonance. 
The frequency dependence of the opacity of a single dipole is illustrated in Fig.~1b.
For a dilute dipolar layer of thickness $\delta z$ and $N$ dipoles per unit volume, the opacity (optical depth) is given by
\begin{align}
\text{OD} = N\sigma \delta z~.
\end{align}
The optical depth increases linearly with $N$, as shown in Fig.~1b. 
However, when the density reaches a value where the dipole spacing $N^{-1/3}\sim \lambdabar$, the resonance is broadened and the efficiency of each dipole is reduced due to dipole--dipole interactions (Fig.~1c-d). This effect is analogous to dipole blockade in Rydberg gases \cite{Lukin2001} but for a thermal gas of two--level atoms it appears as an effective increase in the linewidth $\Gamma=\Gamma_{0}+\Gamma_{\text{dd}}$, where $\Gamma_{\text{dd}}=\beta N$ is typically referred to as self-broadening  \cite{Lewis1980,Maki1991} and $\beta$ is the self-broadening coefficient. 
Self--broadening results in a decrease in the resonant cross-section $\sigma_{0} \sim 1/(N\lambdabar)$ (see Supplementary Information). 
In the high density limit the increased scattering due to adding more dipoles is exactly cancelled by the higher damping rate due to resonant dipole--dipole interactions, leading to the surprising result of a constant broadband opacity (see Fig.~1c-d) given by (see Supplementary Information),
\begin{align}
\text{OD} \sim \frac{\delta z}{\lambdabar}~.
\label{eq:OD2}
\end{align}

\begin{figure}[t]
\centering
\includegraphics[width=0.49\textwidth,angle=0]{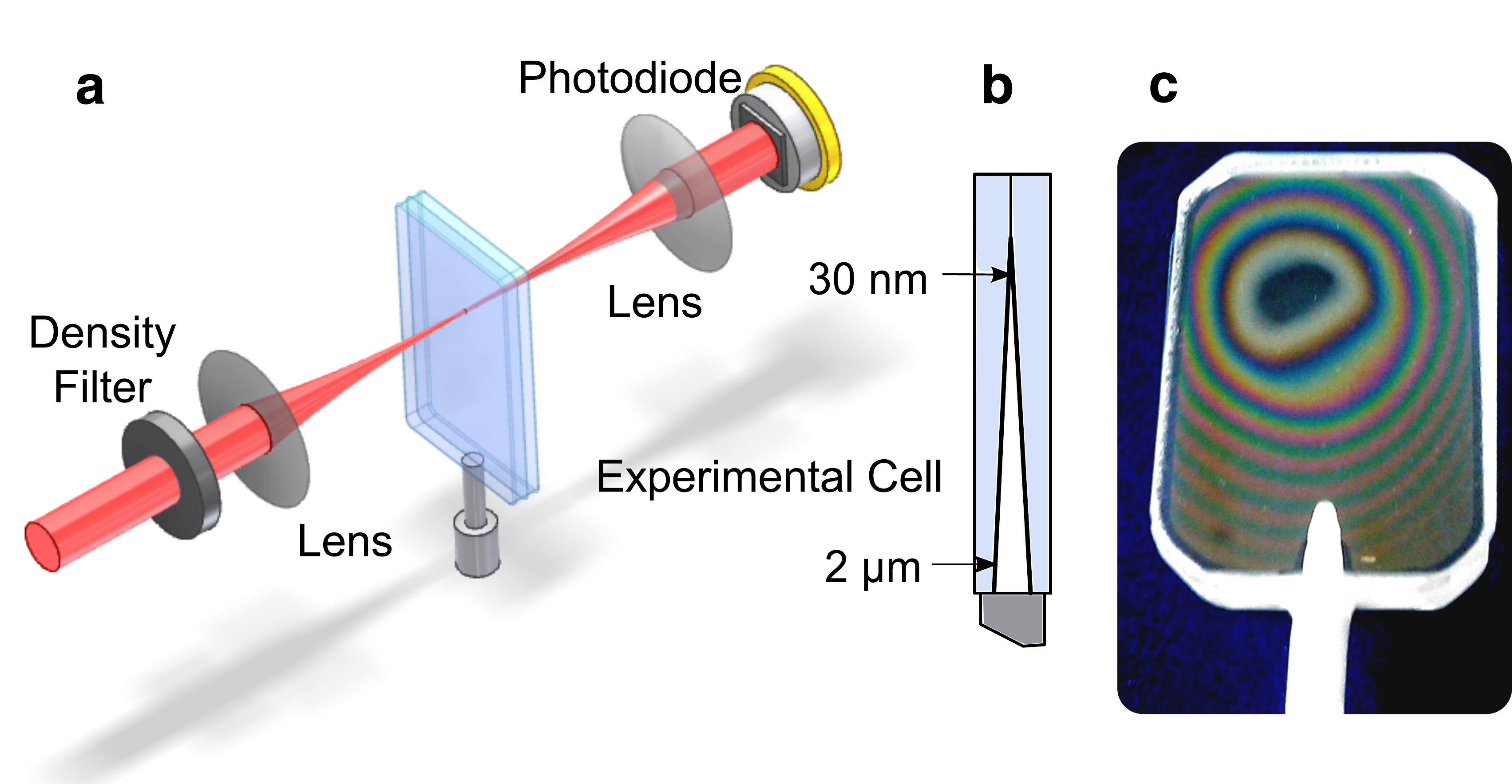}
\caption{Experimental setup and cell.
(\textbf{a}) Schematic of optical setup. A laser is scanned across the D$_{2}$ atomic resonance line in Rb (wavelength 780~nm). The light is filtered to low power to avoid optical pumping effects and focussed to a \unit[30]{$\mu$m} spot size inside the cell, leading to a local thickness variation that is limited by the surface flatness of the windows. The transmission is then recorded on a photodiode. 
	(\textbf{c}) The `Newton's Rings' interference pattern is observed as the gap between the two sapphire windows changes thickness (\textbf{b}), due to the curvature of one of the windows, with a radius $R > 100$~m.
At the centre of these rings the cell has a thickness $\delta z\sim 30$~nm.}
 \label{fig:setup}
\end{figure}
%

As an interesting comparison, if we take a monolayer of atoms (the thickness $\delta z$ is of the order of the Bohr radius, $a_0$) and take the transition wavelength as $\lambdabar \sim a_{0}/\alpha$, where $\alpha$ is the fine structure constant, we obtain an opacity $\text{OD} \sim \alpha$ (see Supplementary Information), very similar to the result of $\pi \alpha$ recently demonstrated for graphene \cite{Nair2008}.

To confirm the saturation of opacity we measure the optical transmission of a single dipolar layer, where the layer thickness must be less than or of the order of $\lambdabar$.
In our experiment we confine a gas of Rb vapour within a nanocell formed by two super--polished sapphire surfaces with a separation between 30~nm and 2~$\mu$m, similar to those in refs. 
\cite{Sarkisyan2001,Sarkisyan2004}. 
To achieve such small separations the plates have a surface flatness of less than 3~nm and a radius of curvature $R>100$~m.
After several iterations of cell design it has been possible to fabricate a nanocell of exceptional quality that allows the scaling of atom--light interactions in a dipolar layer to be elucidated.

The experimental set up and the nanocell are shown in Fig. 2. 
A narrow band laser with a linewidth of less than 1 MHz is frequency tuned in the vicinity of the D$_2$ resonance lines  in Rb ($\lambda = 780$~nm), and the transmission through the cell is recorded at different temperatures. By changing the cell temperature between 20$^{\circ}$C and 350$^{\circ}$C we can vary the number density between $N\lambdabar^{3}\ll1$ and $N\lambdabar^{3}>100$ . In doing so we tune the energy of dipole--dipole interactions ($V_{\rm dd}\propto N$) over 6 orders of magnitude. The measured transmission as a function of laser detuning is shown in Fig. 1d. The similarity with the interacting case in the high density limit (Fig. 1c) is immediately apparent. 
At low temperature the spectrum consists of four main Dicke narrowed resonance lines corresponding to the two isotopes, $^{85}$Rb and $^{87}$Rb, that are both split by the ground state hyperfine interaction (the excited state hyperfine splitting is also resolved in the case of $^{87}$Rb). 
As the number density is increased this structure is completely lost and one observes a broadband absorption that saturates at a finite non--zero value.

To make a more quantative comparision, in Fig. 3 we plot
 transmission spectra for a cell length of $\lambda/2 = 390$~nm together with the predictions of a theoretical model that includes the effect of dipole--dipole interactions. The theoretical calculation is based on a weak probe absorption model \cite{Siddons2008b} including the effects of Dicke narrowing \cite{Romer1955,Briaudeau1998}, cavity effects between the cell walls \cite{Dutier2003a} and dipole--dipole interactions (see \cite{Weller2011a} and references therein). Although this is thicker than a single layer, the saturation of opacity can still be observed.
Additional broadening and shifts due to van der Waals (vdW) atom--surface interactions  are also present for $ \delta z <\lambdabar$ \cite{Sandoghdar1992}. 
We restrict our investigation to $\delta z \gtrsim 90$~nm to avoid vdW atom--surface interactions strongly influencing the resonance lines, such that in the high density limit of interest to this work dipole--dipole broadening dominates.
For a random distribution of two--level dipoles the shift of the resonance (Lorentz shift) is smaller than the broadening \cite{Maki1991} and is neglected.
The effect of the Lorentz shift is apparent in Fig. 3a, but does not change the on resonance opacity significantly.
Overall, Fig. 3 suggests that the model provides an excellent description of the measured transmission.

\begin{figure}[t]
\includegraphics[width=0.47\textwidth,angle=0]{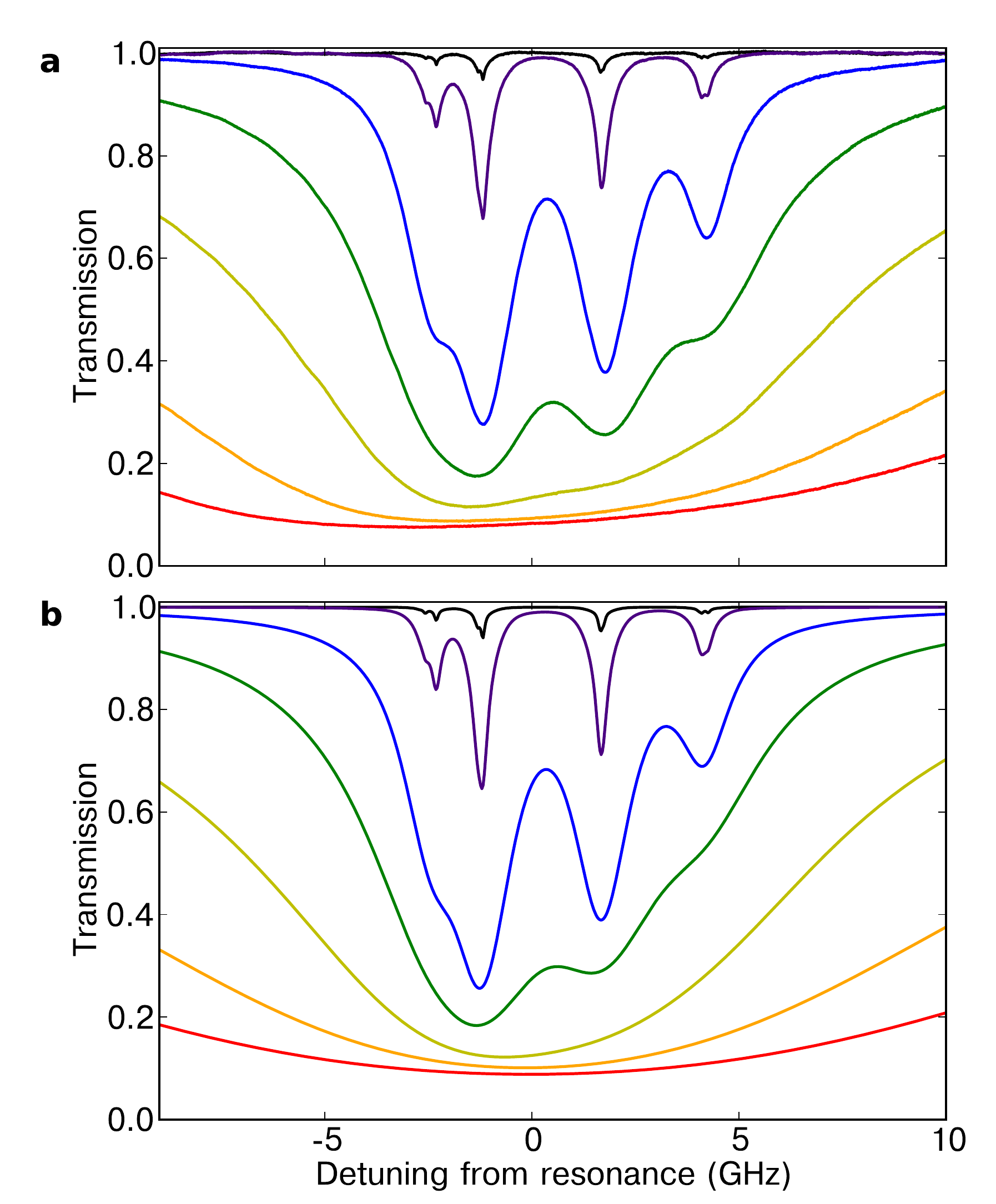}
\caption{Transmission spectra - experiment and theory. 
Experimental (\textbf{a}) and theoretical (\textbf{b}) transmission spectra at thickness $\delta z=390$ nm for measured resevoir temperatures of \unit[130]{$^\circ$C} (black), \unit[160]{$^\circ$C} (purple), \unit[220]{$^\circ$C} (blue), \unit[250]{$^\circ$C} (green), \unit[280]{$^\circ$C} (yellow), \unit[310]{$^\circ$C} (orange), and \unit[330]{$^\circ$C} (red). Theory spectra are based on fitting the experimental data, using a Marquardt--Levenberg method (see \cite{Hughes2010}). Zero on the detuning axis represents the weighted line centre of the D2 line.}
\label{fig:3}
\end{figure}

\begin{figure}[t]
\centering
 \includegraphics[width=0.47\textwidth,angle=0]{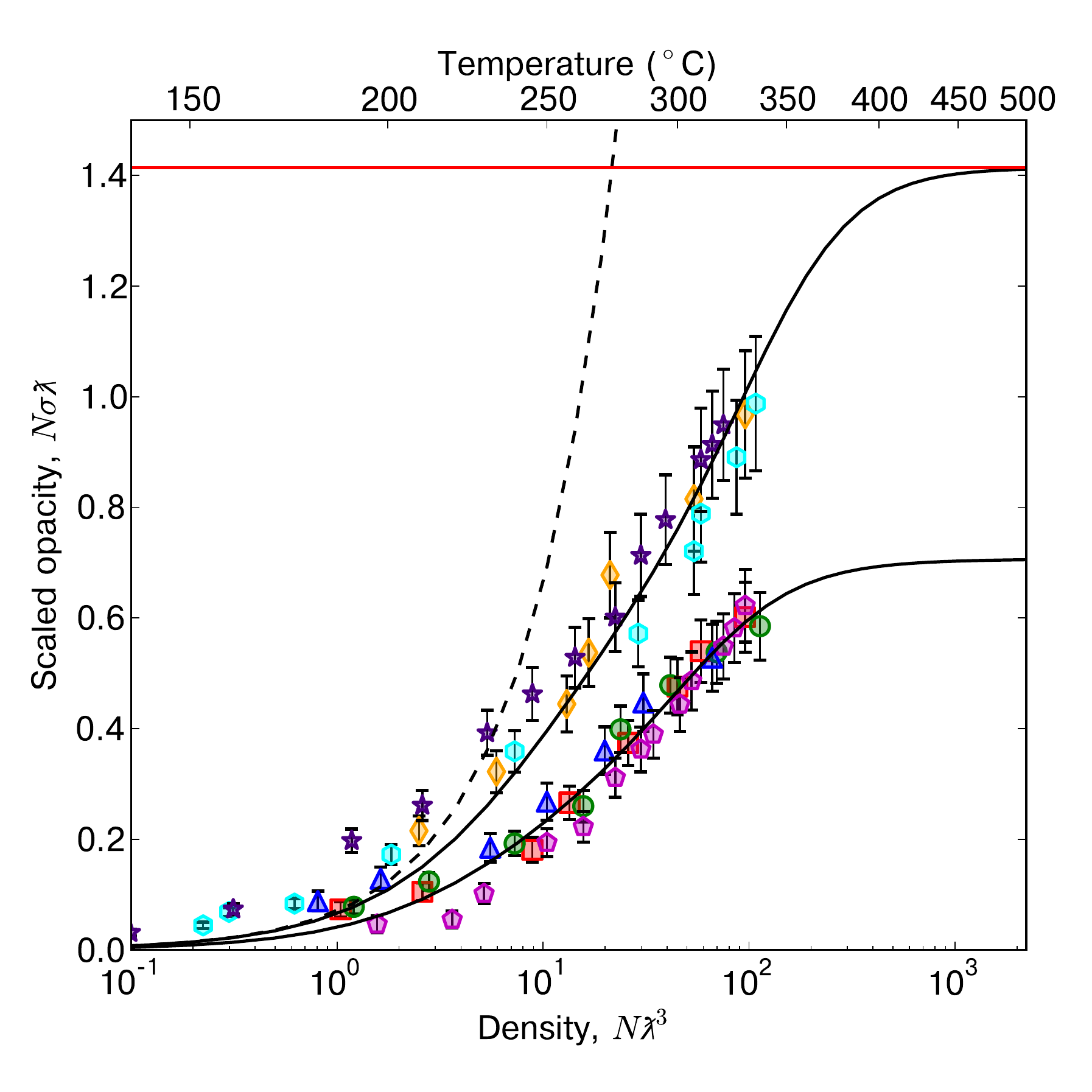}
\caption{Saturation of opacity due to dipole--dipole interactions.
Without dipole--dipole interactions (dashed line) the opacity increases linearly with density. 
When interactions are included into the standard theory (upper solid line) the scaled opacity for the D$_{2}$ line is predicted to saturate at a value $N\sigma\lambdabar=\sqrt{2}$.
However for thicknesses $\delta z < \lambda/4$, we observe a change in the behaviour which doubles the broadening coefficient (lower solid line).  
A constant additional broadening of $\Gamma_{\text{ad}}=2\pi \times \unit[200]{MHz}$ has been added to account for atom--surface effects, which does not affect the value at which the opacity saturates at high density.
Experimental data: $\delta z = 90$~nm (magneta pentagons); $\delta z=110$~nm (red squares); $\delta z=140$~nm (green circles); $\delta z=180$~nm (blue triangles); $\delta z=220$~nm (yellow diamonds); $\delta z=250$~nm (cyan hexagons); $\delta z=390$~nm (purple stars).}
 \label{fig:od}
\end{figure}
%

\begin{figure}[ht]
\centering
 \includegraphics[width=0.46\textwidth,angle=0]{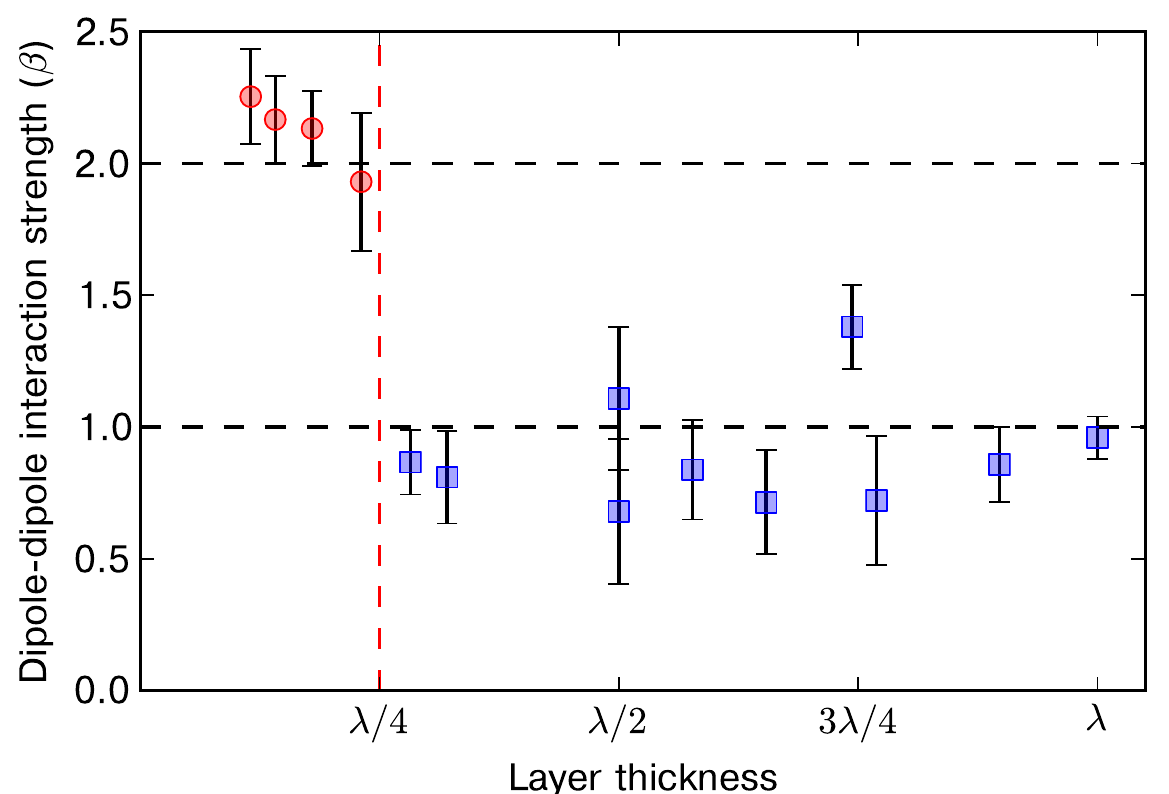}
\caption{Dipole--dipole interaction strength. Interaction strength $\beta$ is plotted as a function of layer thickness. Values and errors are calculated from fitting data of the form shown in Fig. \ref{fig:od} using a Marquardt-Levenberg method. We calculate the average interaction strength for $\delta z < \lambda/4$ (red circles) to be $(2.2 \pm 0.2)\beta$, whereas for $\delta z > \lambda/4$ (blue squares), we find $(0.9 \pm 0.1)\beta$.
}
\label{fig:5}
\end{figure}

To illustrate the saturation effect, we  plot in Fig. \ref{fig:od} the scaled opacity $N\sigma \lambdabar=\text{Im}[\chi]$, where $\chi$ is the susceptibility of the medium (see Supplementary Information), as a function of number density. We take the opacity at the unshifted resonance position of the $^{85}$Rb $F=3\rightarrow F'=4$ hyperfine transition, though in principle one can choose any detuning and achieve similar results.
For layer thicknesses $\lambda/4 < \delta z < \lambda$, the data match our theoretical model including interactions extremely well. Also shown is the theoretical prediction without interactions, where the opacity increases linearly with density and does not saturate. 
For the D2 line, the saturating value is $N\sigma \lambdabar=\sqrt{2}$ following standard theory (see Supplementary Information). 
However for thicknesses $\delta z < \lambda/4$ we observe a deviation from this theory consistent with a dipole--dipole interaction strength $\beta$ that is twice the expected value \cite{Weller2011a}. This arises because for $\delta z < \lambda/4$ atoms moving in opposite directions no longer dephase along the light propagation axis (see Supplementary Information). Remarkably there is a sharp transition between the two regimes at $\delta z=\lambda/4=195$~nm, as shown in Fig. \ref{fig:5} where we plot the effective strength of the averaged dipole--dipole interaction, scaled in units of the self-broadening parameter $\beta$.




In summary, we have demonstrated that adding more scatterers in a thin layer does not make the medium more opaque. Instead the opacity saturates at a well defined value, when dipole--dipole interactions within the medium dominate the light--matter interactions. In addition, when the layer thickness is less than $\lambda/4$, we observe a dramatic change in behaviour whereby the dipolar interaction doubles in strength, resulting in a different opacity that saturates at a value that is half that of the conventional case. This effect is independent of the transition dipole moment, and so is predicted to occur in other systems of two-level interacting dipoles. This result could have important consequences for the generation of non-classical light in confined geometries and in future work we will investigate the photon statistics associated with the transition between a 2D and 3D dipolar layer.

We would like to thank S. J. Clark and M. P. A. Jones for stimulating discussions. We acknowledge financial support from EPSRC and Durham University.



{\scriptsize

%
%
%
%
%
}

\end{document}